# RF phase effect in the ion-guide laser ion source (IG-LIS)

Ruohong Li[1,2,3*], Maryam Mostamand[1,4], Jens Lassen[1,4,5]

[1]TRIUMF - Canada's Particle Accelerator Centre, Vancouver BC, Canada, V6T 2A3
[2]Department of Physics, University of Windsor, Windsor ON, Canada, N9B 3P4
[3]Department of Astronomy & Physics, Saint Mary's University, Halifax NS, Canada, B3H 3C3
[4]Department of Physics & Astronomy, University of Manitoba, Winnipeg MB, Canada, R3T 2N2
[5]Department of Physics, Simon Fraser University, Burnaby BC, Canada, V5A 1S6

**Abstract**

The effect of the phase between the radio frequency (RF) waveform driving the ion guide and the laser pulses generating ions on the intensity of the transmitted ion beam has been studied. Experiments were conducted at TRIUMF's offline laser ion source test stand (LIS-stand) and online at the isotope separator and accelerator (ISAC) facility for radioactive ion beam delivery. In this study, a master clock is used to synchronize the laser trigger for laser ionization and the RF waveform generator driving the ion guide, so that laser ionization within the ionization volume inside the RF ion guide will occur at a specific RF phase which affects the ions' transmission through the RFQ. At optimal phase the ion extraction from the IG-LIS can be improved by 10-50%. Simulations were run considering both fringe field and RFQ phase effects. The method also provides an additional function of IG-LIS to modulate the laser-ionized ions at hundreds of kHz, allowing phase-sensitive detection for experiments downstream. In addition, modulation of the RF envelope (on and off like a gating device) in transmission mode allows for the suppression of surface-ionized species outside the laser-ion pulse, which provides an alternative to a classic fast kicker for beam purification.

## 1. Introduction

Resonant ionization laser ion sources (RILIS), combining high ionization efficiency and elemental selectivity, are widely used at isotope separator online (ISOL) facilities [1,2]. However, traditional hot-cavity RILIS suffers from isobar contamination from surface-ionized species. To decouple the laser ionization region from the hot target and transfer tube, the radio frequency quadrupole (RFQ) ion-guide laser ion source was proposed and put into use at TRIUMF [3], Mainz University [4], CERN-ISOLDE[4], and elsewhere. In this device, the laser ionization of neutral atoms can be generated in a cold environment and the RF quadrupole field on a linear ion guide confines and guides the ions to the extraction region, while surface ionized species are retained in the hot target and transfer tube region by use of a repeller electrode or biasing the target and transfer tube. Isobar suppression factors of up to $10^6$ have been achieved. This device, called ion-guide laser ion source (IG-LIS) at TRIUMF or laser ion source and trap (LIST) at Mainz and ISOLDE, is routinely used at both ISOL facilities to provide highly purified beams for experiments.

Further improvements and developments are ongoing with a focus on increasing efficiency and incorporating additional functionalities. A segmented IG-LIS was proposed as a beam-bunching device [5,6]. By sending the excitation laser perpendicularly to the effusing atom beam, high-resolution in-source laser spectroscopy was conducted using LIST [7]. SIMION simulations were performed to find optimal IG-LIS operation parameters for isobar suppression [8]. The simulation results showed: 1) replacing the quadrupole with an octupole can reduce the energy spread; 2) an increase in laser repetition rate from 10 kHz to 50 kHz can increase the laser ionization efficiency of IG-LIS at suppression mode by up to 3×.

---

[*] corresponding email: ruohong@triumf.ca

In this work, the effect of the RF waveform phase relative to the laser ionization timing is investigated on IG-LIS. Since the ion beam traveling through the RFQ has a finite diameter and momentum distribution, the RFQ with periodically changing voltages on the rods will have varying acceptance in phase space, which affects the ion transmission. P. Dawson [9] calculated the percentage of ion transmission of a linear RFQ as a function of the beam diameter and initial RF phase (the phase at the moment the ion enters the RFQ). D. Lefaivire et al. [10] experimentally observed the phase effect on the RFQ ion transmission using a continuous beam of $N_2^+$ ions. When they used a bunched krypton ion beam, a six-fold increase in the difference between the minimum and maximum transmission was observed as a function of the initial phase. J. Lavoie et al. [6] discussed the application of this effect for the segmented IG-LIS design.

For RILIS, excitation and ionization lasers are pulsed lasers typically with 10 kHz repetition rate and 10-50 ns pulse length (~10 ns for dye lasers and ~50 ns for Ti:Sa lasers). Since the frequency of the RF applied to IG-LIS is typically 0.5-1.5 MHz, the RF period is much longer than the laser pulses. The excitation rate of atomic transitions is typically $10^7$-$10^8$ s$^{-1}$ for the first step and $10^5$-$10^6$ s$^{-1}$ for the following steps. Therefore, laser resonance ionization takes place within at most a hundred nanoseconds. If the laser-ionized ions are generated inside the transfer tube, due to the thermal velocity distribution at the birth the ion pulse length will be stretched before entering the IG-LIS. The thermal velocity distribution of atoms is determined by the temperature of the transfer tube and the element mass. Assuming the transfer tube is at 2200 °C, the element is heavy, and the transfer tube is long, the ion pulse can span several RF cycles [11] when arriving at the entrance of the ion guide. In that case, the phase effects will be averaged across the ion pulse and hence do not affect the total transmission. However, these ions generated in the hot environment before the IG-LIS are typically "containment ions" to be suppressed/repelled when the IG-LIS is operated in isobar suppression mode. The ions laser-ionized within the ion guide's cold ionization volume are from the neutral atoms effusing out of the hot transfer tube. They are generated in that cold region within a hundred nanoseconds, with all ions experiencing a specific RF phase at the moment of their creation. Depending on how the suppression mode is set, either by biasing the target and transfer tube to a lower potential or by applying a higher potential on the repeller electrode, there is always some distance between the potential barrier and the RFQ entrance. Ions generated in this transition region will experience the RF phase effect not only in the radial direction but also in the axial direction due to the fringe field at the RFQ entrance. This portion of ions cannot be neglected in suppression mode since most ions are generated at the beginning of the RFQ due to the divergence of the atom beam effusing from the transfer tube. Even for the ions generated inside the hot transfer tube, the effect can be observable when the ion pulse is short, which can be possible if the tube temperature is low, the element is light, and the transfer tube is short (for TRIUMF online target and sources, the IG-LIS transfer tube is ~10 mm long vs. the usual 60 mm long standard RILIS transfer tube). In normal IG-LIS operation, the phase effect is averaged out when the RF driver and laser trigger are not synchronized. To study this effect and investigate its role in enhancing IG-LIS efficiency, a master clock was used to synchronize the RF waveform generator and the laser trigger generator.

## 2. Offline experimental test using laser-ionized Pr

The experimental test was conducted at TRIUMF's offline laser ion source test stand (LIS-stand) [12]. The Pr sample (50 μl, Alfa Aesar Specpure atomic spectroscopy standard solution, 1 μg/μl Pr in the form of Pr(NO$_3$)$_x$ in 2% HNO$_3$) was pipetted onto a thin foil and dried at 110 °C in an oven. The sample-loaded Ti foil was folded and placed inside the 15 mm long 3 mm diameter Ta crucible (similar to the transfer tube in the online target and source module) and installed in the LIS-stand. After pumping down the vacuum chamber below 5×10$^{-6}$ Torr the crucible was resistively heated up to 1600 °C. The Pr atoms, created by the heating and chemical reduction process inside the crucible, were two-step resonantly excited and ionized by the pulsed lasers at 491.539 nm + 833.401 nm [12] (all wavelengths in the paper are vacuum wavelengths). The generated ions were then guided through a 50 mm long RFQ [3], which is the standard IG-LIS module used at TRIUMF for online beam delivery. At the end of the RFQ, the ions were extracted and accelerated to 10 keV and then deflected 90° vertically away from the incident laser beam path. After the electrostatic bender, the ions will be focused and directed into a deceleration region and a quadrupole mass spectrometer (ABB Extrel MAX-300) for mass analysis and ion counting. With the ion beam energy decelerated below 50 eV a typical mass resolution $\Delta m \sim 1$ amu can be achieved.

The laser beams were generated in the adjacent laser laboratory using two birefringent-filter (BRF) tuned Ti:Sa lasers. The Ti:Sa lasers were simultaneously pumped by a frequency-doubled Nd:YAG laser with 10 kHz repetition rate and 20 W output power. The pulse length of the Ti:Sa lasers is about 50 ns (FWFH). The laser power was 450 mW for the first step transition at 491.539 nm, and 1.3 W for the second step transition at 833.401 nm. Both laser beams were expanded 2-4 times, overlapped via dichromic mirrors, and focused with a $f = 5$ m uncoated 2-inch diameter lens into the IG-LIS and crucible inside the vacuum chamber of the LIS-stand 5 m away. To efficiently ionize the Pr atoms inside the source, the laser pulses need to be superposed in both space and time. The pump laser of the two Ti:Sa lasers was triggered by a digital delay gate generator (Berkeley Nucleonics BNC577). The timing difference between the two Ti:Sa lasers, caused by varying gain at different wavelengths, was adjusted using intra-cavity Pockels cells. The laser wavelengths were constantly monitored by a wavelength meter (HighFinesse WS6-200) multiplexed with a 4-channel fiber switcher.

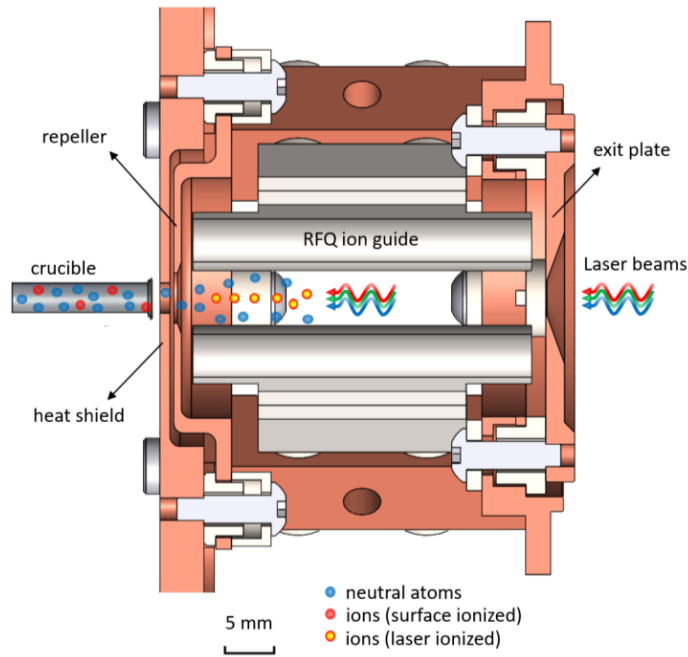

Fig. 1. Cross-sectional view of the IG-LIS used at the LIS-stand. The ion guide's free-field radius $r_0$ is 5 mm and the RFQ rod radius is 3 mm. The length of the IG-LIS rod is 34 mm. The offline crucible has 3 mm inner diameter and 15 mm length.

To examine the phase effect on the ion transmission inside IG-LIS, the RF waveform generator (Keysight/Agilent 33511B) and laser-trigger delay gate generator (BNC 575) were synchronized using the 10 MHz reference clock from the BNC575. As presented in our previous work [3] the TRIUMF IG-LIS is driven by a 50% duty cycle square waveform, instead of using a sinusoidal RF. The latest revision RF driver uses SiN transistors to allow RF up to 3 MHz and ±100 V amplitude, which gives the ion guide a high transmission for singly charged ions from 2 to 240 amu. The cross-sectional view of the IG-LIS with scale is shown in Fig.1.

The square waveform (TTL) is sent to the RF driver transistor switch module inside the high-voltage cage via an optical fiber. Inside the cage, two DC voltage power supplies with opposite polarity are connected to the driver to provide the RF voltage. The driver output flip-flops between the output voltages of two DC power supplies. The driver output is sent to the IG-LIS via a Kapton insulated twisted pair transmission line, which is shielded with silver-coated steel braid to further reduce RF leakage and noise pickup. The length of the twisted pair transmission line between the RF driver and IG-LIS is about 6 meters for the LIS-stand. The open-end transmission line impedance is 68 Ω. The RFQ charge/discharge time is about 120 ns [3,13].

To measure the time profile of the extracted ion beam a multi-channel analyzer (FAST ComTec MCA4) was used. The measurements were triggered by the laser trigger at 10 kHz to measure the time profile of the ions within the 100 μs laser-pulse period. For each measurement, 29078 scans were averaged.

## 3. Phase effect in IG-LIS transmission and suppression mode

First, the laser frequencies and alignment were optimized for Pr laser ionization with unsynchronized clocks for the laser trigger and RFQ waveform generator. The IG-LIS was set at transmission mode with the crucible potential at 5 V and all other electrodes – namely the repeller, RFQ bias, and exit electrode - all at 0 V. In this paper, the electrode potential values are relative to the source high voltage (HV) potential. The RF amplitude $V_{rf}$ (0 to peak amplitude) was optimized to 45 V at 0.4 MHz to get the maximum ion signal. Fig. 2a shows the results with the laser on and off. The ion pulse length is long with the majority of ions emerging within 45 μs and the wing spanning over 100 μs (the laser repetition period).

Fig. 2b shows the result with the laser trigger pulse and RFQ waveform synchronized and phase-locked. The effect of the initial RF phase on ion transmission is prominently observed on both laser-ionized ions (laser on) and surface-ionized ions (laser off) as a periodic modulation of ion signal amplitude. The period ~2.5 μs matches the RF frequency $f$ = 0.4 MHz. In Fig. 2c and 2d, we gated the RF by applying an envelope modulation on the RF signal. The envelope modulation signal is at 10 kHz repetition rate to match the laser pulse repetition. The "RF-on" window can be adjusted based on the timing and width of the laser-ion time profile. In suppression mode, there is no need to use this gating technique to suppress the surface ionized ions. Gating the RF amplitude is useful in transmission mode to clean up the beam as an alternative to downstream beam purification with a fast beam kicker. Using a 30 μs gating window, up to 70% of the ions generated within the hot crucible - primarily surface-ionized ions - can be suppressed. One notable feature apparent in Fig. 2c and 2d, distinguishing from Fig. 2a, is the presence of $2f$ component. To understand these observed features, a numerical simulation of the phase effect of the IG-LIS has been conducted.

Using Dawson's theoretical method [9] we can calculate the ion transmission as a function of the initial phase, i.e. the RF phase that an ion experiences when entering the RFQ. The result is shown in Fig. 3. The incident ion positions in both the x and y directions are assumed to follow a Gaussian distribution centered along the RFQ z-axis, with standard deviations σ = 1 mm, 2 mm, and 3 mm. The velocities at both x and y directions are randomly chosen within the Maxwell-Boltzmann distribution by applying the thermal temperature of the crucible and the mass of Pr atoms. The RF waveform is assumed as $V_{rf} \sin(\omega t + \varphi)$ without loss of generality. As presented in Dawson's paper [9] and shown in Fig. 3, the smaller the incoming beam diameter, the greater the relative transmission and the smaller the phase effect. It is worth noting, that this theoretical calculation uses the field potential of hyperbolic quadrupole rods, which is slightly different from that of our circular rods.

In the calculation, the ions start at the beginning of the RFQ rods with isotropic position and velocity distribution in both the x and y directions. That is why the transmission peaks at both 0- and 180-degrees phases (Fig. 3), as the x and y directions are indistinguishable for the ions. However, this phase effect disagrees with our experimental result shown in Fig. 2b, where the transmission varies with the same frequency as the RF frequency $f$, instead of $2f$. A similar $1f$ transmission variation was also observed in Lefaivre's work [10] where a low-resolution mass filter was investigated with a continuous $N^{2+}$ ion beam.

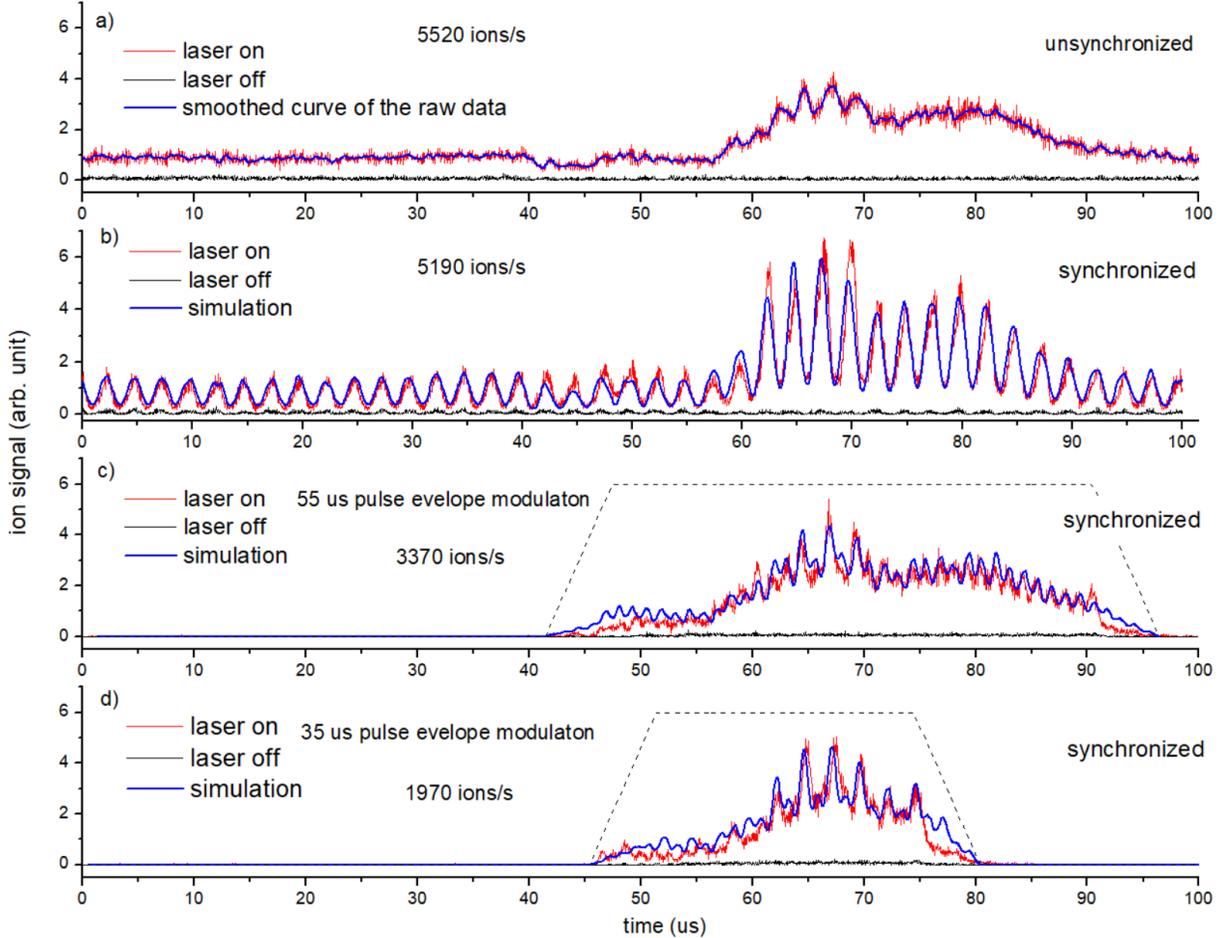

Fig. 2. Time profile of Pr ions in **transmission mode** (crucible bias = +5 V, all other electrodes at the source potential). a): the laser trigger and RFQ waveform generator were not synchronized; b)-d): the laser trigger and RFQ waveform generator were synchronized; c) and d): the RF waveform was envelope-modulated by a 55 and 35 μs RF-on window. The red and black lines are the experimental data for laser on and off, respectively and the blue lines are theoretical simulations. The simulation parameters: b) a = 1, b = 0.6, and $\varphi$ = 2.3 rad using eq. (1) model; c) a = 0.3, b = 0.05, c = 1.0, $\varphi$ = 2.3 rad, and $\varphi'$ = 3.1 rad using eq. (2) model; d) a = 0.25, b = 0.15, c = 1.0, $\varphi$ = 2.3 rad, and $\varphi'$ = 2.6 rad using eq. (2) model. The total ion rates in a) b) c) d) conditions are noted on the plots, respectively. The statistical uncertainty of the total ion rates is about ±150 ions/s.

In Lefaivre's experiment, the $N^{2+}$ ions were all generated before the RFQ. In this work, ions can also be generated before the repeller (majority in the transmission mode), between the repeller and the RFQ, inside the RFQ, and at the end of the RFQ (in both transmission and suppression mode). Except for the regular phase effect calculated in Fig. 3 ($2f$ variation with the initial phase), there is a fringe field between the repeller and the RFQ (~2 mm), which particularly affects the fast ions that fly through this region within a few RF cycles, as well as the ions generated close to the RFQ entrance. In this region, the fringe field has components in both radial and axial directions, resulting in different variations in ion transmission compared to the regular phase effect. Holme et al.[14] also observed a similar $1f$ variation in transmission with a high-resolution mass filter. In their work, they experimentally demonstrated that the $1f$ variation was caused by the fringe field. They measured the dependence of the ion transmission variation with time upon the number of RF cycles an ion experiences in the fringe field region. It shows the variation is only significant for ions that experience fewer than 4 RF cycles in the fringe field. The dependence is complex and only agrees to a limited extent with the theoretical calculation of Dawson [15]. In our case, the Pr atoms are generated in the crucible at 1600 ℃, the ion mean speed is ~530 m/s, which corresponds to 1.5 cycles of 0.4 MHz RF for ions traversing the 2 mm gap between the repeller electrode and the RFQ. Our IG-LIS operation conditions fall in the region where the fringe-field phase effect is critical.

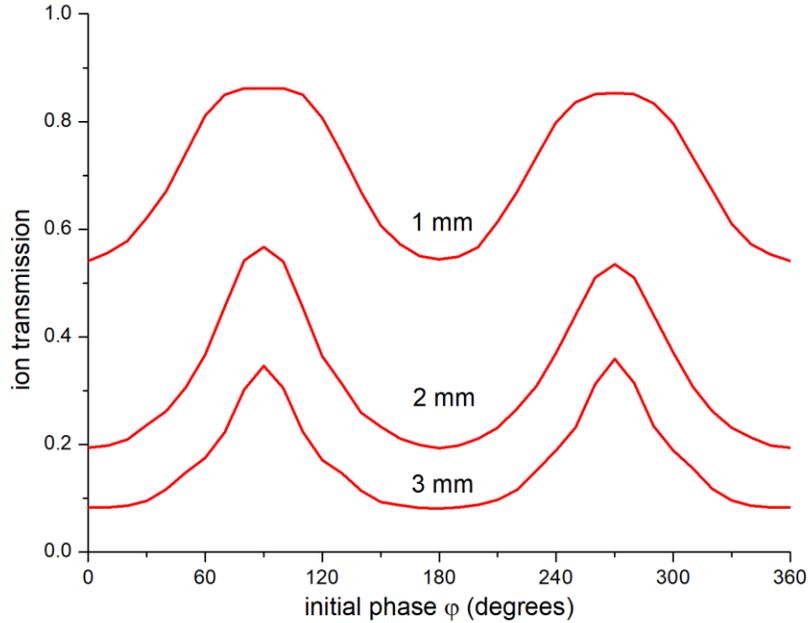

Fig. 3. Plot of simulated ion transmission through the RFQ of the IG-LIS (without considering fringe field) as a function of the initial phase (in degree) for the beam sizes with standard deviation σ = 1mm, 2 mm, and 3 mm. The RF waveform is assumed as $V_{\mathrm{rf}}\sin(\omega t + \varphi)$, where φ is the initial phase and $V_{\mathrm{rf}} = 45$ V. The mass was set as 141 amu for Pr.

Holme's experiment [14] also showed that the practical significance of the fringe-field phase effect is only limited to high-resolution RFQ filters. However, Lefaivre's experimental results [10], along with the findings of this work, provided a counterexample. Dawson explained this discrepancy [16] could be caused by the quadrupole size relative to the incident ion beam size. Lefaivre's RFQ was hyperbolic and had free-field radius $r_0$ = 20 mm; Holme's RFQ had $r_0$ = 2.76 mm. The incoming ion beam sizes unfortunately were not given in the references. However, based on the $r_0$ dimensions given, very likely the ratio of incident ion-beam radius to $r_0$ in Lefaivre's experiment was smaller as compared to Holme's. Based on Dawson's calculations the fringe-field phase effect increases as the relative ion beam size decreases, which is opposite to the regular phase effect since the axial-direction electric field plays a significant role, as shown in Fig. 11 of Ref. [9]. Therefore, Holme's conclusion based on their practical experiments might fail when the relative dimensions of the RFQs change. Our experiment also shows a strong fringe-field phase effect (Fig. 2b) even though the IG-LIS was used as an ion guide with zero mass resolution. In our IG-LIS, $r_0$ = 5 mm and the incoming ion beam diameter is 1-3 mm.

As demonstrated by Dawson [16], the dependence of ion transmission on the fringe field is complex. Dawson's theoretical calculations showed only limited agreement with experimental results, partly due to the fragmented and limited nature of the experimental data available for comparison. In this work, due to the good regularity of the $1f$ amplitude variation on ion transmission (Fig. 2b), we use a highly simplified model to simulate the effect:

$$P_{\text{syn}} = P_{\text{un-syn}} * (a + b\sin(\omega t + \varphi)) \qquad (1)$$

Here ω is the RF angular frequency (2π*0.4 MHz) and φ is the initial phase; $P_{\text{syn}}$ is the time profile of ion transmission when the laser trigger and RF generator were synchronized (phase locked) and $P_{\text{un-syn}}$ is that when unsynchronized. In the simulations, the smoothed data (blue in Fig. 2a) from the raw data (red in Fig. 2a) was used for $P_{\text{un-syn}}$. The first term with coefficient "a" represents the ions unaffected by the fringe-field phase effect. The second term with coefficient "b" represents the ions affected by the initial phase condition. As shown in Fig. 2b, the theoretical model result (blue) matches the experimental data (red) well with the parameters a = 1, b = 0.6, and $\varphi$ = 2.3 rad, except for a slight discrepancy in the 50-70 μs region.

In the simple theoretical model here, all ions are treated as identical without distinguishing their initial conditions (position, potential, and velocity distribution), which is a rough approximation. As studied both theoretically and experimentally before[11,17], the different parts of the ion time profile are generated at different regions of the laser ion source (LIS). This manifests as a double-bump structure with the first bump occurring at 56-73 μs and the second at 73-90 μs shown in Fig. 2a. For hot-cavity LIS, it is typically interpreted as: the ions in the first bump are generated inside the hot cavity and those in the second one are from the region inside the transfer tube [11]. The typical structure sometimes has a narrow and small leading peak before the double bumps, which is not prominent in our result. Those ions are generated at the exit region of the hot cavity, where the extraction field significantly penetrates [11]. This interpretation was derived from theoretical simulations based on the conditions of hot-cavity LIS, where the hot cavity and transfer tube are at the same potential. However, in the case of IG-LIS, the electrodes - namely the target/crucible, repeller electrode, RFQ, and exit electrode - can be at different potentials. In the transmission mode shown in Fig. 2 (crucible bias = 5 V, all other electrodes in 0 V), the first bump stems from ions generated near the entrance of the RFQ, where the neutral atom density is high. The second bump, only visible in transmission mode, is from ions generated inside the crucible (transfer tube in the online module). There is a wide plateau (45-95 μs) beneath the two main peaks. This signal stems from ions generated inside the central part of the RFQ [17]. Because of the length of the RFQ, it spans widely. The experimental results for the IG-LIS phase effect in suppression mode are shown in Fig. 4 with the repeller potential at 5 V, while all other IG-LIS electrodes are at 0 V. In the suppression mode, the ions from the crucible will be repelled and the second bump observed in transmission mode disappears. Therefore, the plateau at 45-95 μs can be seen clearly (Fig. 4a).

The substructures (a series of small peaks) within the first bump in both Fig. 2a and Fig. 4a are difficult to interpret. It might be due to the complex electrode structure of the IG-LIS, therefore complex potential distribution near the entrance of the RFQ (heat shield, repeller, and RFQ entrance), similar to the substructure observed in the GdB$_6$ hot cavity LIS [16].

In Fig. 2c and 2d, the $2f$ regular phase effect appears. The RF amplitude gating causes the absence of the RF field as ions pass through the fringe-field region, which significantly reduces the fringe-field effect. As a result, the regular phase effect, previously overshadowed by the fringe-field effect as shown in Fig. 2b, becomes apparent. To model this new feature, the simulation has been modified from eq. (1) to include an extra phase-dependent term:

$$P_{\text{syn}} = G * P_{\text{un-syn}} * \left(a + b\sin(\omega t + \varphi) + c * P_{\text{phase}}(\varphi')\right) \qquad (2)$$

Here $P_{phase}$ is the numerically simulated result of the regular phase effect in Fig. 3, with the parameters $V_{rf}$ = 45V and σ = 1mm; $G$ is the gate pulse profile, which is equal to 1 during the gate open period (35 and 55 μs for Fig. 2c and 2d situations, respectively) and equal to 0 when the gate is off. The rise/fall edge of the gate pulse was estimated to be ~ 6 us, which has been included in the profile of $G$. The simulation results based on Eq. (2) model are plotted in Figs. 2c and 2d, showing a good match with most of the structures observed in the experimental data, though not in all regions.

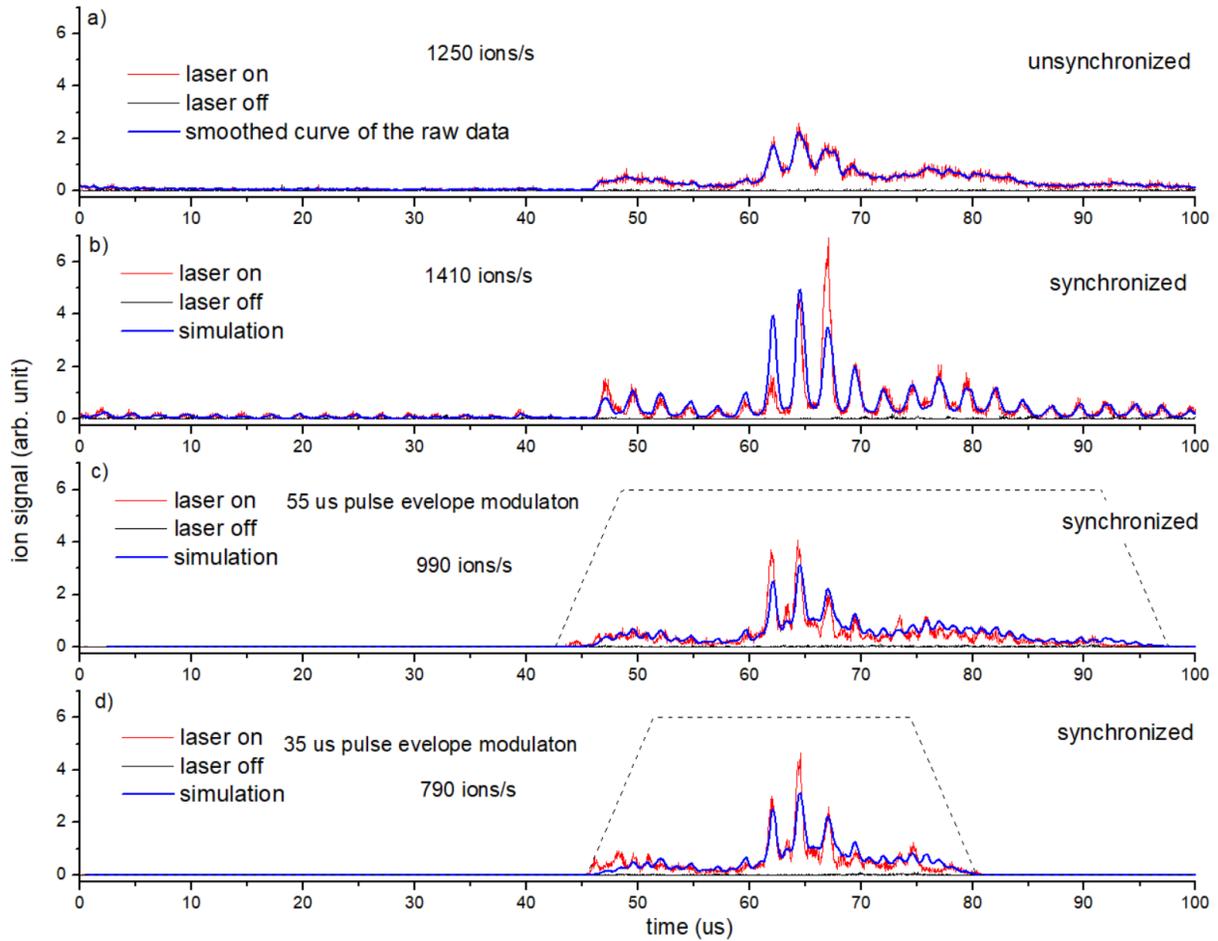

Fig. 4 Time profile of Pr ions in **suppression mode** (repeller = +5 V, all other parts at the source potential, which means the detected ions are only generated after the repeller electrode and within the RF ion guide). a): the laser trigger and RFQ waveform generator were not synchronized; b)-d): the laser trigger and RFQ waveform generator were synchronized; c) and d): the RF waveform was envelope modulated by a 55 and 35 μs RF-on window (shown as the black dash lines). The red and black lines are the experimental data for laser on and off, respectively and the blue lines are theoretical simulations. The simulation parameters using eq. (2) are: b): a = 0.25, b = 0.9, c = 1.3, $\varphi$ = 2.6 rad and $\varphi'$ = 2.5 rad; c) and d): a = 0.0, b = 0.15, c = 1.5, $\varphi$ = 2.6 rad and $\varphi'$ = 2.5 rad. The total ion rates in a) b) c) d) conditions are noted on the plots, respectively. The statistical uncertainty of the total ion rates is about ±150 ions/s.

This is reasonable, as the simplified model of Eq. (2) cannot account for all aspects of complexity:

1) the different parts of the time profile are generated in distinct regions of the IG-LIS, and therefore have different ion birth environments and velocity distributions, which cannot be fully described by Eq. (2);

2) $P_{phase}$ represents the transmission rate if the ions started at the RFQ entrance and passed through the entire length of the RFQ. In reality however, a fair portion of ions are generated inside the RFQ, which has slightly higher transmission than what Fig. 3 plots;

3) at the end of the RFQ, the extraction field penetrates into the RFQ. This subtly is not captured in the model expressed in eq. (2).

Additionally, eq. (2) only considers the three groups of ions: one unaffected by the phase condition, one primarily affected by the fringe field, and another primarily affected by the regular phase effect. However, some ions are likely influenced by both effects with comparable weight. This higher-order term is neglected in the model.

The experimental results above indicate significant phase effects on the time profile of transmitted ions. However, these effects may be averaged out and become unobservable in the total transmission rate for high-mass elements, such as $^{141}$Pr, which has a long pulse length of ~50 μs. The averaged ion count rates in different operation modes and experimental conditions were measured and presented in Fig. 2 and Fig. 4. Being synchronized or not, the total transmission rate is essentially unchanged in transmission mode (Fig. 2a and 2b), but increases by about 10% in suppression mode when properly synchronized (Fig. 4a and 4b). Fig. 4b shows the results after phase optimization, achieved by adjusting the initial phase of the waveform on the RF generator. Before this optimization, the ion rate was 7% lower.

## 4. Offline experimental test using Ag

To further study the phase effect, more experimental tests of IG-LIS were done on different elements. In the offline test stand, an IG-LIS phase test on Ag was conducted. The ionization scheme for Ag was 328.163 nm + 827.576 nm + 532 nm (non-resonant ionization). The laser powers were 20 mW for 328.163 nm, 15 mW for 827.576 nm, and 6.7 W for 532 nm. The Ag ions generated and transmitted through the IG-LIS as a function of the initial phase of the RF waveform were measured (shown in Fig. 5). Here, the initial phase of the RF waveform refers to the initial phase at which the waveform is generated by the generator, which can be set on the generator.

The top right plot in Fig. 5 is in transmission mode, while all other plots are under suppression mode, either with the repeller voltage > 0 V or the source bias < 0 V. Two transmission valleys are shown within $2\pi$ period, with one higher and one lower. This is due to the combination of both fringe-field and regular phase effects. Even in the transmission mode (Fig. 5b top plot), this $2f$ phase effect is evident, which did not show significantly in the Pr experiment (Fig. 2b). This implies a reduced fringe-field effect in the Ag case.

The averaged ion rate over a $2\pi$ period is represented by the blue dashed line in Fig. 5, which expectedly aligns with the ion rate measured using unsynchronized clocks (blue solid line). By setting the RF waveform phase to the optimized point, the extracted ion rate from the IG-LIS can be improved by 20-50% in suppression mode. Experimentally, the optimal phase value varies depending on several parameters. It changes slightly with the potential of the ion source and the repeller, but more significantly with the RF frequency.

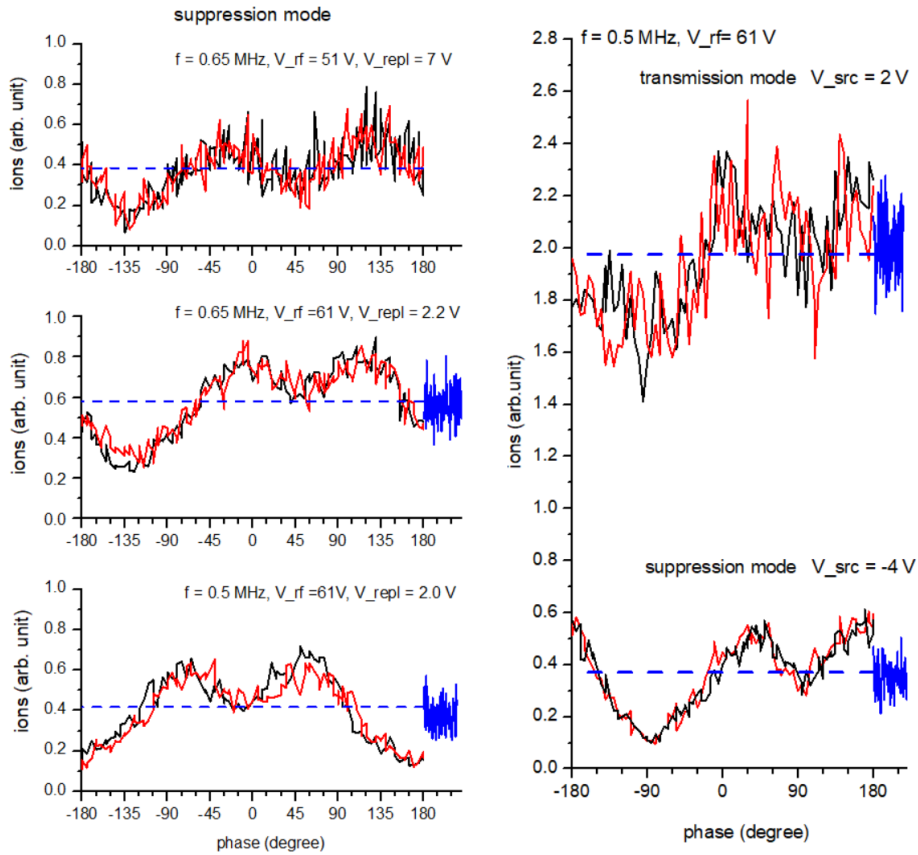

Fig. 5 The total transmitted Ag ions as a function of the RF waveform phase (the set value on the RF generator). The black (scanning up the phase) and red solid lines (scanning down the phase) are the experimental IG-LIS ion rates as a function of the phase. The straight blue dashed line is the average value of the ion rate over a full $2\pi$ period, which matches the result measured with unsynchronized clocks (blue solid line). Note the unsynchronized data was not measured for the top plot of a). The repeller (repl) and source (src) potentials are noted in the plots. Except noted, all other electrodes of the IG-LIS are at the source potential. Left: all in suppression mode with different repeller potentials and RF paramters. Right: comparison of the phase effect in transmission mode ($V_{src}$=2V) and suppression mode ($V_{src}$=-4V).

## 5. Experimental test online using U

The IG-LIS phase effect has also been studied online at ISAC using $^{238}$U. $^{238}$U was extracted from a UCx target irradiated by 9.8 µA of 480 MeV protons. The target heating current was 340 A and the transfer line was resistively heated with a current of 270 A (standard ISAC target ion source operating parameters). U was laser ionized using two-step resonant excitations to an autoionization (AI) state using 436.328 nm + 367.840 nm laser light. The time profiles of the laser-ionized U ions are shown in Fig. 6. They were taken from a channel electron multiplier downstream of the ISAC high-resolution mass separator, with the beam attenuated by a factor of 100. The applied RF had $V_{rf}$ = 37 V with a frequency of 0.4 MHz. The repeller potential was at +5 V, and all other IG-LIS electrodes were at 0 V.

An expected sinusoidal-like modulation of transmitted ion rate was observed at the initial phase of the RF waveform from -90 to 180 degrees. However, an abrupt increase in ion rates occurred at -120 degrees, remained high at -150 degrees, and returned to a reasonably expected value at -180 degrees (which should be roughly the same value as 180 degrees). We currently have no explanation for this anomalous behavior. From the time profiles, an additional group of ions suddenly appeared in the 75-80 μs region when the initial phase was set to -120 degrees. Limited by the online beam time, this issue could not be thoroughly investigated. Nevertheless, based on the observed change of ion rates corresponding to the phase, a 20% ion rate improvement was achieved by optimizing the phase.

The amplitude modulation in the U time profile is visible but not distinct mainly due to the lower beam intensity online. Additionally, the high mass of $^{238}$U, which is approximately double that of $^{107}$Ag and $^{141}$Pr, reduces the influence of both the fringe field and the regular phase effects. Conversely, these effects should be more pronounced in lighter elements such as Mg or Be. Further experiments on the IG-LIS phase effect are planned using lighter isotopes for comparison.

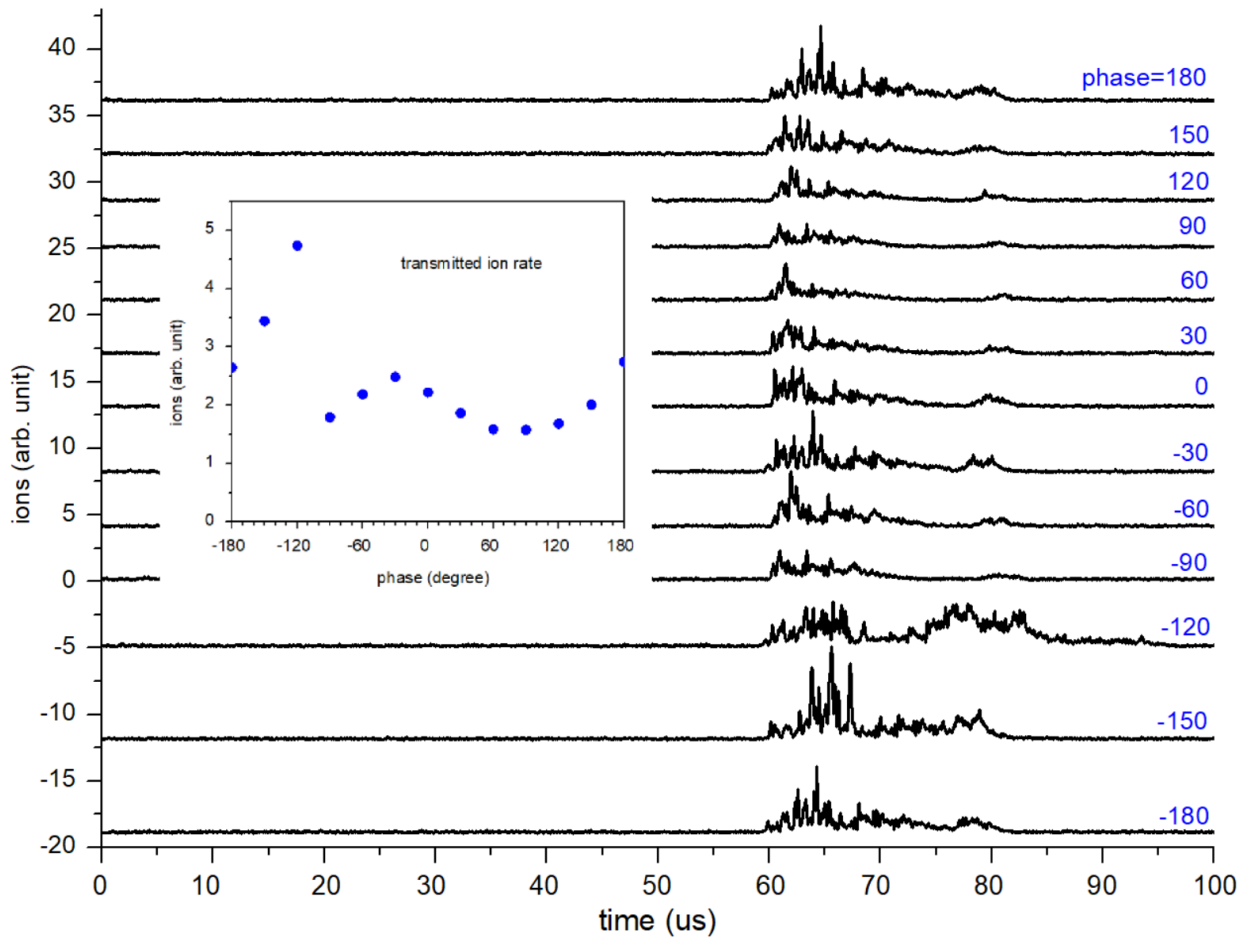

Fig. 6 Time profiles of laser ionized U at different RF waveform phases (the set values on the RF generator), measured online at ISAC with 9.8 μA, 480 MeV proton beam on a UCx target with IG-LIS. The inset plot shows the variation of total transmitted ion beam intensity as a function of the initial phase of the RF waveform. The RF frequency was at 0.4 MHz and $V_{rf} = 37$ V. The IG-LIS was set at **suppression mode** with the repeller potential at +5 V and all other electrodes at 0 V. The data traces are offset for clarity and are presented together in a single plot. For original data, the baselines of all traces are around zero.

## 6. Summary

In this work, the phase effect of IG-LIS has been studied offline at LIS-stand and online at the ISAC facility. In the experiments with laser-ionized Pr, Ag, and U, the ion rates from the IG-LIS were improved by 10-50% in suppression mode by synchronizing the laser trigger and RF waveform and optimizing the RF phase. We plan to extend these investigations into the light mass region in the future. Theoretical simulations were performed to understand the $1f$ and $2f$ amplitude modulation in the time profiles of the transmitted ions and to explore how the initial phase of the RF affects ion transmission.

With the option to modulate the envelope of the RF, we can swiftly switch the RF on and off to gate the time profile of the ion beam in transmission mode (Fig. 2c and 2d). This gives an alternate means of reducing surface-ionized isobars outside of the laser ion pulse length by a factor of 2-3. The $1f$ modulation in the amplitude of transmitted ions, induced by the fringe-field phase effect, exhibits high regularity and can be effectively described by the simple sinusoidal function in Eq. (1). This opens the possibility for phase-sensitive detection in collinear laser spectroscopy (CLS) to improve the signal-to-noise ratio. The test and implementation of this technique are planned at the LIS-stand and ISAC polarizer beamline[18].

**Acknowledgments**

The experimental work is funded by TRIUMF which receives federal funding via a contribution agreement with the National Research Council of Canada and through the Natural Sciences and Engineering Research Council of Canada (NSERC) Discovery Grant RGP-IN-2021-02993 to R. Li and RGP-IN-2024-659 to J. Lassen. M. Mostamand acknowledges funding through the University of Manitoba graduate fellowship.

**Reference**

[1] V. Fedosseev, K. Chrysalidis, T.D. Goodacre, B. Marsh, S. Rothe, C. Seiffert, and K. Wendt, "Ion beam production and study of radioactive isotopes with the laser ion source at ISOLDE," J. Phys. G Nucl. Part. Phys. **44**(8), 084006 (2017).

[2] J. Lassen, R. Li, S. Raeder, X. Zhao, T. Dekker, H. Heggen, P. Kunz, C.D. P. Levy, M. Mostanmand, A. Teigelhöfer, and F. Ames, "Current developments with TRIUMF's titanium-sapphire laser based resonance ionization laser ion source: An overview," Hyperfine Interact. **238**(1), 33 (2017).

[3] S. Raeder, H. Heggen, J. Lassen, F. Ames, D. Bishop, P. Bricault, P. Kunz, A. Mjøs, and A. Teigelhöfer, "An ion guide laser ion source for isobar-suppressed rare isotope beams," Rev. Sci. Instrum. **85**(3), 033309 (2014).

[4] F. Schwellnus, K. Blaum, R. Catherall, B. Crepieux, V. Fedosseev, T. Gottwald, H.-J. Kluge, B. Marsh, C. Mattolat, S. Rothe, T. Stora, and K. Wendt, "The laser ion source trap for highest isobaric selectivity in online exotic isotope production," Rev. Sci. Instrum. **81**(2), 02A515 (2010).

[5] K. Blaum, C. Geppert, H.-J. Kluge, M. Mukherjee, S. Schwarz, and K. Wendt, "A novel scheme for a highly selective laser ion source," Nucl. Instrum. Methods Phys. Res. Sect. B Beam Interact. Mater. At. **204**, 331–335 (2003).

[6] J.P. Lavoie, P. Bricault, J. Lassen, and M.R. Pearson, "Segmented linear radiofrequency quadrupole/laser ion source project at TRIUMF," Hyperfine Interact. **174**(1–3), 33–39 (2007).

[7] R. Heinke, T. Kron, S. Raeder, T. Reich, P. Schönberg, M. Trümper, C. Weichhold, and K. Wendt, "High-resolution in-source laser spectroscopy in perpendicular geometry: Development and application of the PI-LIST," Hyperfine Interact. **238**(1), 6 (2017).

[8] M. Mostamand, R. Li, J. Romans, F. Ames, P. Kunz, A. Mjøs, and J. Lassen, "Production of clean rare isotope beams at TRIUMF ion guide laser ion source," Hyperfine Interact. **241**(1), 36 (2020).

[9] P.H. Dawson, "A detailed study of the quadrupole mass filter," Int. J. Mass Spectrom. Ion Phys. **14**(4), 317–337 (1974).

[10] D. Lefaivre, and P. Marmet, "Optimizing ion injection phase in quadrupole mass filters," Rev. Sci. Instrum. **45**(9), 1134–1137 (1974).

[11] Y. Liu, C. Baktash, J.R. Beene, C.C. Havener, H.F. Krause, D.R. Schultz, D.W. Stracener, C.R. Vane, Ch. Geppert, T. Kessler, K. Wies, and K. Wendt, "Time profiles of ions produced in a hot-cavity resonant ionization laser ion source," Nucl. Instrum. Methods Phys. Res. Sect. B Beam Interact. Mater. At. **269**(23), 2771–2780 (2011).

[12] R. Li, J. Lassen, P. Kunz, M. Mostamand, B.B. Reich, A. Teigelhöfer, H. Yan, and F. Ames, "Lu and Pr beam development for resonance ionization laser ion sources," Spectrochim. Acta Part B At. Spectrosc. **158**, 105633 (2019).

[13] J.-P. Lavoie, "Production of pure ion beams by laser ionization and a fast release RFQ," PhD Thesis, Université Laval (2010).

[14] A.E. Holme, and W.J. Thatcher, "The dependence of ion transmission on the initial r.f. phase of a quadrupole mass filter," Int. J. Mass Spectrom. Ion Phys. **10**(3), 271–277 (1973).

[15] P.H. Dawson, "Fringing fields in the quadrupole mass filter," Int. J. Mass Spectrom. Ion Phys. **6**(1–2), 33–44 (1971).

[16] P.H. Dawson, "Fringing fields and other imperfections," in *Quadrupole Mass Spectrom. Its Appl.*, (Elsevier, 1976), pp. 95–119.

[17] D.A. Fink, S.D. Richter, K. Blaum, R. Catherall, B. Crepieux, V.N. Fedosseev, A. Gottberg, T. Kron, B.A. Marsh, C. Mattolat, S. Raeder, R.E. Rossel, S. Rothe, F. Schwellnus, M.D. Seliverstov, M. Sjödin, T. Stora, P. Suominen, and K.D.A. Wendt, "On-line implementation and first operation of the Laser Ion Source and Trap at ISOLDE/CERN," Nucl. Instrum. Methods Phys. Res. Sect. B Beam Interact. Mater. At. **344**, 83–95 (2015).

[18] R. Li, J. Lassen, C.D.P. Levy, M. Roman, A. Teigelhöfer, V. Karner, G.D. Morris, M. Stachura, and A. Gottberg, "Recent upgrades and developments at TRIUMF's laser nuclear-spin-polarization facility," Nucl. Instrum. Methods Phys. Res. Sect. B Beam Interact. Mater. At. **541**, 228–231 (2023).